\documentclass[%
 reprint,
 amsmath,amssymb,
 aps,
]{revtex4-1}

\newcommand{\Ncoll}{\mbox{$N_{\mathrm{coll}}$}}
\newcommand{\Npart}{\mbox{$N_{\mathrm{part}}$}}
\newcommand{\pp}{$pp$}
\newcommand{\pPb}{$p$+Pb}
\newcommand{\pA}{$p$+A}

\newcommand{\dAu}{$d$+Au}
\newcommand{\NN}{$NN$}
\newcommand{\sigmaNN}{\mbox{$\sigma_{\mathrm{NN}}$}}
\newcommand{\ETj}{\mbox{$E_{\mathrm{T}}^j$}}
\newcommand{\ETpj}{\mbox{$E_{\mathrm{T}}^{j(p)}$}}
\newcommand{\ETevt}{\mbox{$E_{\mathrm{T}}^{\mathrm{evt}(p)}$}}
\newcommand{\kevt}{\mbox{$k^\mathrm{evt}$}}
\newcommand{\ET}{\mbox{$E_{\mathrm{T}}$}}
\newcommand{\ETmin}{\mbox{$E_{\mathrm{T}}^{\mathrm{min}}$}}
\newcommand{\ETmax}{\mbox{$E_{\mathrm{T}}^{\mathrm{max}}$}}

\newcommand{\SigmaET}{\mbox{$\Sigma{E}_{\mathrm{T}}$}}
\newcommand{\pT}{\mbox{$p_{\mathrm{T}}$}}
\newcommand{\RpA}{\mbox{$R_{p\mathrm{A}}$}}
\newcommand{\TpA}{\mbox{$T_{p\mathrm{A}}$}}

\usepackage{graphicx}
\usepackage{dcolumn}
\usepackage{bm}
\usepackage{hyperref}

\usepackage{lineno}
\usepackage{tabularx}

\begin{document}



\title{Calculation of centrality bias factors in $p$+A collisions based on a positive correlation of hard process yields with underlying event activity}

\author{Dennis V. Perepelitsa}\email{dvp@bnl.gov}
\author{Peter A. Steinberg}%
 \email{peter.steinberg@bnl.gov}
\affiliation{%
 Brookhaven National Laboratory, Upton, New York 11973, USA
}%

\date{\today}

\begin{abstract}
Hard scattering yields in centrality-selected proton-- and deuteron--nucleus (\pA) collisions are generally compared to nucleon--nucleon (\NN) cross-sections scaled to the appropriate partonic luminosity using geometric models derived from an analysis of minimum bias \pA\ interactions. 
In general, these models assume that hard process rates and the magnitude of the soft event activity in the underlying \NN\ collisions is uncorrelated. 
When included, these correlations influence the measured yield in a nominal centrality interval, an effect typically referred to as a ``centrality bias''. 
In this work, the impact of a positive correlation between the hard scattering yield and the underlying event activity in individual \NN\ collisions is investigated.
This correlation is incorporated into the centrality calculations used by ATLAS and PHENIX, both based on a similar Monte Carlo Glauber approach but with different models of the per-collision or per-participant event activity. 
It is found that the presence of this correlation tends to increase the yield measured in more central events and decrease it in peripheral events.  
Numerical factors to correct measured yields for the centrality bias effect are calculated for \pA\ collisions at RHIC and the LHC.
Reasonable agreement with a previous calculation of these factors by PHENIX is found, despite differences in the implementation of the underlying bias.
\end{abstract}

\maketitle

\section{Introduction} 

High transverse momentum (\pT) probes of proton- and deuteron--nucleus (\pA) collisions serve a number of purposes within high energy nuclear physics~\cite{Salgado:2011wc,Albacete:2013ei}. They offer access to the underlying partonic content of the projectile and target, and may thus measure the modification of nucleonic parton densities in the presence of the nucleus~\cite{Accardi:2004be,Eskola:2009uj}. Furthermore, they may be sensitive to the effects of multiple soft scatterings inside the nucleus, which may manifest as the energy loss of partons before the hard scattering or the angular decorrelation of the outgoing partons~\cite{Kang:2012kc,Xing:2012ii,Kutak:2012rf}. In certain kinematic regimes dominated by small Bjorken-$x$ in the nucleus, they may even test models of non-linear QCD evolution or other novel effects~\cite{Gelis:2010nm}.

Rates of hard probes such as charged particles or jets are especially important to explore as a function of the collision geometry, which may constrain the impact parameter dependence of parton density modifications~\cite{Helenius:2012wd} or probe saturation phenomena that vary with the local nuclear density~\cite{Rezaeian:2012ye,Albacete:2012xq,Tribedy:2011aa}. More generally, \pA\ collisions serve as an overall test of the relationship between the rate of hard processes, arising from the point-like scatterings of nucleon constituents~\cite{RevModPhys.59.465}, and the total soft particle activity, driven by the successive interactions of the proton with the nucleons in the target nucleus~\cite{PhysRevLett.41.285}.

Hard process rates are typically reported as a function of the \pA\ event {\em centrality}, an experimental classification of the collision geometry generally based on a measurement of the underlying event (UE) activity in a rapidity region separated from the hard process of interest. Geometric parameters such as the mean number of nucleon--nucleon (\NN) collisions (\Ncoll) or total number of nucleon participants (\Npart) for each centrality interval are estimated using the distribution of UE activity in minimum bias \pA\ collisions. Thus, the question naturally arises as to whether the rare events which produce a high-\pT\ charged particle or jet have the same relationship between geometric quantities and centrality as do minimum bias events. In fact, \NN\ collisions with a hard scattering are observed to be accompanied by a larger magnitude of transverse energy or charged particle multiplicity~\cite{PhysRevD.65.092002,Khachatryan:2010pv,Aad:2010fh} in the underlying event. However, this fact is not typically included in centrality-dependent measurements of the hard scattering yield, resulting in a so called ``centrality bias''~\cite{Adler:2006xd} that must be systematically corrected.

In measurements of inclusive particle and jet yields in centrality-selected \pA\ collisions, the per-event yields of a particular hard process, $\mathrm{d}N/\mathrm{d}\pT$, are tested against the expectation from an incoherent superposition of individual \NN\ collisions. Deviations from this hypothesis are quantified through the nuclear modification factor $\RpA \equiv \left. \mathrm{d}N/\mathrm{d}\pT \right/ \TpA \mathrm{d}\sigma^{pp}/\mathrm{d}\pT$, where \TpA\ is the mean value of the nuclear thickness function for a given centrality selection and $\mathrm{d}\sigma^{pp}/\mathrm{d}\pT$ is the hard probe cross-section in \pp\ collisions. In a Monte Carlo (MC) Glauber approach, $\TpA$ is related to \Ncoll\ via $\TpA \equiv \left<\Ncoll\right> / \sigmaNN$, where $\sigmaNN$ is the inelastic \NN\ cross-section. Correcting for a centrality bias in the hard scattering yield has been explored by the PHENIX Collaboration~\cite{Adare:2013nff}, where it was modeled as an increase in the mean of the specific UE multiplicity distribution associated with the \NN\ collision producing a hard scattering. After this correction, measurements of the yields of identified hadrons~\cite{Adler:2006wg,Adare:2013esx} in deuteron--gold (\dAu) collisions at RHIC show collision scaling ($\RpA = 1$ in all centrality intervals) at intermediate \pT\  ($3$--$10$~GeV) and mid-rapidity.

In preliminary measurements of the centrality-selected hadron yields in proton--lead (\pPb) collisions at the LHC by the ATLAS Collaboration~\cite{ATLAS-CONF-2013-107}, no such correction is yet applied. Thus deviations of $10$--$20$\% from the geometric expectation at intermediate hadron \pT\ ($3$--$20$~GeV) have been observed, typically resulting in an $\RpA > 1$ in the most central, or high activity, events and $< 1$ in the most peripheral, or low activity, events. Similarly, preliminary measurements of $Z$ boson yields~\cite{ATLAS-CONF-2014-020} found that collision scaling only holds after the application of a simple correction for the centrality bias effect. 

On the other hand, preliminary measurements of very high-\pT\ jets in \dAu~\cite{Perepelitsa:2013jua} and \pPb~\cite{ATLAS-CONF-2014-024} collisions have unexpectedly reported the opposite modification pattern, in which $\RpA < 1$ ($>1$) in central (peripheral) events. These modifications are thought to be associated with large Bjorken-$x$ in the projectile, $x_p > 0.1$, which may give rise to the observed effect due to the associated proton configurations interacting more weakly than average with the nucleons in the target nucleus~\cite{Alvioli:2014eda} or with the exclusion of these partons from QCD evolution~\cite{Bzdak:2014rca}. Thus, a quantitative understanding of possible centrality biases is needed to better characterize the modifications.

In this paper, we present a model to estimate the size of the centrality bias effect arising from a positive correlation between the UE activity and average hard scattering yield in individual \NN\ collisions. We posit that for \NN\ events which produce some total UE multiplicity or transverse energy (referred to in the subsequent discussions as \ET\ for simplicity), the average yield of final state objects $Y_\mathrm{hard}$ (which may be high-\pT\ hadrons, jets, electroweak bosons, etc.) produced through hard scattering rises linearly with \ET,

\begin{equation}
\left<Y_\mathrm{hard}( E_{T} )\right> \propto E_{T}.
\label{eq:intro}
\end{equation}

This relationship can be understood schematically through a geometric picture of \pp\ collisions, in which the hard scattering rate and the magnitude of UE activity both depend on the extent of the transverse \pp\ overlap region. Thus, the UE activity and the yield are intercorrelated through a mutual correlation with the impact parameter of the collision. This hypothesis has been considered before in, for example, Refs.~\cite{Frankfurt:2010ea,Jia:2009mq} and has also been described in terms of multiple parton--parton interactions~\cite{Sjostrand:1986ep,Sjostrand:1987su}. Recently, the rates of $J/\psi$~\cite{Abelev:2012rz} and $\Upsilon$~\cite{Chatrchyan:2013nza} production in \pp\ collisions at the LHC have been observed to be proportional to the soft particle multiplicity, in agreement with this hypothesis.

Additionally, Eq.~\ref{eq:intro} was motivated by MC studies using event generators tuned to soft observables at the LHC. Namely, $10^6$ minimum bias Pythia 8.183~\cite{Sjostrand:2007gs} \pp\ events, incorporating the leading order MSTW2008 parton distribution function set~\cite{Martin:2009iq} and tuned to measurements of minimum bias observables by ATLAS in \pp\ collisions~\cite{ATLAS:2012uec}, were generated for $\sqrt{s} = 2.76$~TeV and $5.02$~TeV. In the latter case, the system was also boosted by $\Delta{y} = +0.465$ to match the \NN\ kinematics in the recent \pPb\ data-taking at the LHC.

\begin{figure}[!t]
\includegraphics[width=\columnwidth]{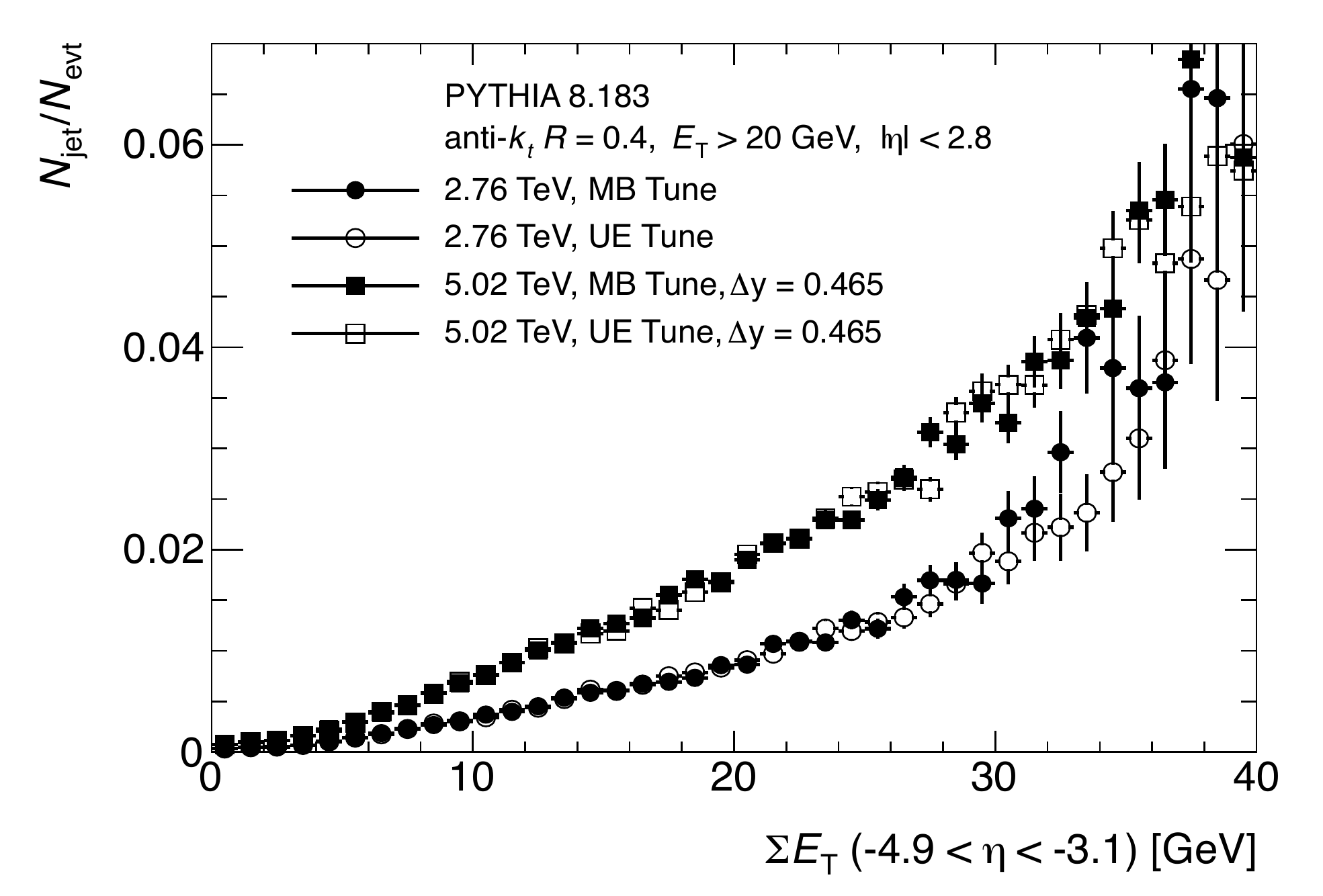}
\caption{\label{fig:PYTHIA} Per event yield of jets with $\pT > 20$~GeV and $\left|\eta\right| < 2.8$, plotted as a function of the underlying event \SigmaET. Results are shown for Pythia 8 simulations of \pp\ $2.76$~TeV collisions (circles) and for \pp\ $5.02$~TeV collisions with a $\Delta{y} = +0.465$ rapidity shift with respect to the lab frame (squares) where the \SigmaET\ is measured in $-4.9 < \eta < -3.1$. }
\end{figure}

The sum of the transverse energy, $\Sigma\ET$, of all final-state, visible particles was measured within $-4.9 < \eta < -3.1$ to match the acceptance of the forward calorimeter in the ATLAS experiment. Furthermore, jet reconstruction with an $R=0.4$ anti-$k_\mathrm{t}$ algorithm~\cite{Cacciari:2008gp} was run on the final-state, visible particles. The mean per-event yield of jets with $\pT > 20$~GeV and $\left|\eta\right| < 2.8$ is shown as a function of $\Sigma\ET$ in Fig.~\ref{fig:PYTHIA} at both $\sqrt{s}$ energies. The yield was found to be approximately linear in \SigmaET, with the largest possible deviations only in the high-$\Sigma\ET$ ($>30$ GeV) tail. This quantitative relationship persisted when the generator was tuned instead to measurements of the UE in the presence of a high-\pT\ track or cluster~\cite{ATLAS:2012uec}.

\section{Model Overview} 

This section describes the procedure for determining the effects of the correlation in \NN\ collisions described by Eq.~\ref{eq:intro} on measurements of the centrality-selected yield in \pA\ collisions. 

Three models are considered: in the Uncorrelated Model (UCM), there is no correlation in individual \NN\ collisions between the magnitude of the UE activity \ET\ and the hard scattering yield $Y$; in the Partially Correlated Model (PCM), the two are related according to Eq.~\ref{eq:intro}, with an additional stochastic term accounting for the randomization of the impact parameter of the proton in each successive $p$+$N$ collision; in the Variably Correlated Model (VCM), the stochastic term is integrated out and the strength of the correlation is controlled by a single parameter.

In these models, a \pA\ collision at a given \Npart\ is treated as a superposition of individual \NN\ collisions, which on average contribute equally to the total UE activity measured in the detector (called here the centrality signal) and to the total yield. The specific hard process could be, for example, the yield of high-\pT\ reconstructed jets at mid-rapidity.

The distribution produced in the centrality detector arising from each nucleonic participant $j$, \ETj, is modeled by a Gamma distribution, $\Gamma(\ETj; k, \theta)$, characterized by shape parameter $k$ and scale parameter $\theta$. Thus, the probability to produce some \ETj\ in any \NN\ collision $j$ is

\begin{eqnarray}
P_{NN}( \ETj ) & = &  \Gamma(\ETj; k, \theta) \nonumber \\
& \equiv & \left. \left(\ETj\right)^{k-1} \exp\left(-\ETj/\theta\right) \right/ \Gamma(k) \theta^k,
\label{eq:basicNNgamma}
\end{eqnarray}

\noindent where $\Gamma(k)$ is the standard Gamma function. 

Now consider the set of \pA\ collisions with a fixed value of \Ncoll. In an event with a set of \ETj\ values associated with each nucleonic participant, $\left\{ \ETj \right\}$, let the total centrality signal, \ET, be 

\begin{equation}
\ET \equiv \left(\sum_{j=1}^{\Ncoll} \ETj\right) + \ETevt,
\label{eq:ETsum}
\end{equation}

\noindent where \ETevt\ is a contribution to the total \ET\ from the proton participant, and is drawn from a separate Gamma distribution characterized by parameter \kevt. (In a pure wounded nucleon model, $\kevt = k$, but it is retained as a distinct parameter for now.) Thus, for fixed \Ncoll, the distribution of total $\ET$, $P_{N\mathrm{coll}}(\ET)$, is given by the $\Ncoll$-convolution of $P_{NN}( \ETj )$, convolved with the  contribution from the proton participant, 

\begin{equation}
P_{N\mathrm{coll}}( \ET ) = \Gamma(\ET; \Ncoll  k, \theta ) \oplus \Gamma(\ET; \kevt, \theta).
\label{eq:prob_Ncoll}
\end{equation} 

The total \ET\ distribution for all \pA\ events is obtained by taking the convolution of Eq.~\ref{eq:prob_Ncoll} with the per- \pA\ event probability of \Ncoll\ collisions $P(\Ncoll)$, obtained from a Glauber MC simulation~\cite{Alver:2008aq}. This formulation is appropriate for experimental centrality schemes in which the signal varies monotonically with \Npart. 

\subsection{Uncorrelated Model} Let $Y_j$ be the average per-collision yield in a given \NN\ collision $j$. In the limit of no correlation between UE activity and hard scattering rate, the average yield in any \NN\ collision $j$ is

\begin{equation}
Y_j = C,
\label{eq:yield_WN}
\end{equation}

\noindent where $C$ is the \ETj-independent average yield. For rare hard scattering events $C \ll 1$ but this is not a requirement for the model. The average yield in \pA\ collisions with fixed \Ncoll, $Y_{N\mathrm{coll}}(\ET)$, is 

\begin{equation}
Y_{N\mathrm{coll}}(\ET) = \sum_{j=1}^{\Ncoll} Y_j = C \Ncoll.
\end{equation}

If $Y_{N\mathrm{coll}}$ and \ET\ are uncorrelated, then for fixed \Ncoll\ the yield-weighted and minimum bias \ET\ distributions have the same shape, but with an overall scale difference of $C \Ncoll$. Thus, for any centrality selected ($\ETmin < \ET < \ETmax$) \pA\ events, the estimated $\left<\Ncoll\right>$ and the total yield are simply related by the overall rate $C$. Therefore, in the UCM the yield in each centrality selection scales with the number of binary collisions estimated from analyzing the \ET\ distribution in minimum bias events, and there is no specific centrality bias. 

\subsection{Partially Correlated Model} Now consider the case with an overall correlation in each \NN\ collision between the produced \ETj\ and the average yield $Y_j$, as related in Eq.~\ref{eq:intro}. In this scenario, each participant contributes equally to the centrality signal on average, and the proton is a participant in each \NN\ collision. To harmonize these concepts, let \ETj\ and $\ETpj$ be the contributions to the centrality signal arising from the participating nucleon and proton in each \NN\ collision, respectively, where $\ETpj$ is drawn from the same Gamma distribution as \ETj. Since the impact parameter in the proton is effectively randomized between successive collisions, $\ETpj$ is chosen at random for each \NN\ collision. The average yield in a given \NN\ collision $j$ is a function of the sum of these two terms,

\begin{equation}
Y_j = C \left. \left( \ETj + \ETpj \right) \right/ 2k\theta,
\label{eq:yield_hardsoft}
\end{equation}

\noindent where the $1/2k\theta$ term is a normalization factor corresponding to the mean value of $\ETj + \ETpj$ and keeps the overall normalization $C$ the same as in Eq.~\ref{eq:yield_WN}. 

On the other hand, none of the $\ETpj$ appear in the total centrality signal in Eq.~\ref{eq:ETsum} and an ``event-wide'' proton contribution is added only via the $\ETevt$ term for the entire \pA\ collision. The average yield from \pA\ collisions at a given \Ncoll\ and set of $\left\{ \ETj, \ETpj \right\}$ values is 

\begin{equation}
Y_{N\mathrm{coll}}(\ET) = \frac{C}{2k\theta} \sum_{j=1}^{\Ncoll} \left( \ETj +  \ETpj \right).
\label{eq:yield_hardsoft_all}
\end{equation}

When integrating over all \ET, the total yield in the PCM is the same as in the UCM. However, \pA\ configurations at fixed \Ncoll\ with a large \ET\ will contribute a larger fraction of the total yield. 

\subsection{Variably Correlated Model} If Eq.~\ref{eq:yield_hardsoft} is averaged over many \NN\ configurations, one can integrate out the $\ETpj / k\theta$ term and replace it with its mean value of $1$. In that case, the average hard scattering yield in any \NN\ collision $j$ reduces to

\begin{equation}
Y_j = C \left. \left( \left.\ETj \right/ k\theta + 1 \right) \right/ 2.
\label{eq:yield_hardsoft2}
\end{equation}

As will be seen later, even though this replacement removes the stochastic component of the correlation between the yield and the UE activity in \NN\ collisions, the resulting centrality bias is numerically identical within statistical precision. This suggests that in addition to the PCM, the magnitude of the bias can be explored as a continuous function of the strength of the correlation between the hard scattering yield and UE activity. The correlation strength is specified by a parameter $0 < \alpha < 1$ such that the average yield in any \NN\ collision $j$ is

\begin{eqnarray}
Y_j = C \left( \alpha  \left.\ETj \right/ k\theta + (1 - \alpha) \right),
\label{eq:yield_hardsoft3}
\end{eqnarray}

\noindent and the average yield at fixed \Ncoll\ is therefore

\begin{eqnarray}
 Y_{N\mathrm{coll}}(\ET)
 = C \left( \alpha\smash{\sum_{j=1}^{\Ncoll}} \ETj / k \theta + ( 1 - \alpha ) \Ncoll \right). 
\label{eq:yield_hardsoft4}
\end{eqnarray}

Numerically, the choice of $\alpha=\frac{1}{2}$ is equivalent to the (\ETpj-integrated) PCM, while $\alpha=0$ describes the UCM.

\begin{figure}[!t]
\includegraphics[width=\columnwidth]{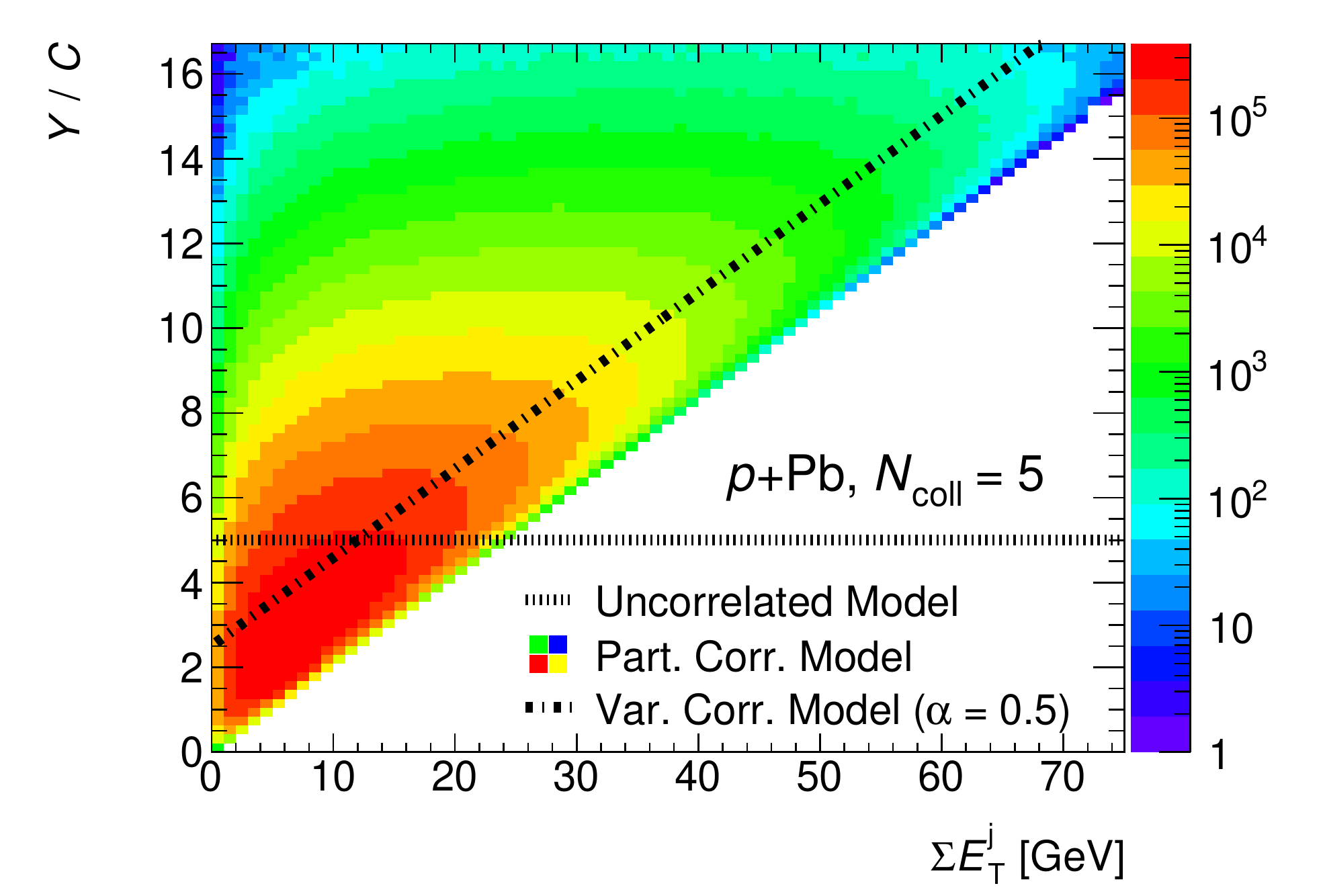}
\caption{\label{fig:example_corr} Correlation between the normalized yield $Y/C$ and the centrality signal arising from the nucleon participants in the nucleus, $\sum\ETj$, shown for \pPb\ events with $\Ncoll = 5$ for the UCM (horizontal line), PCM (colored squares) and VCM with $\alpha=0.5$ (diagonal line). The contribution from the proton participant, \ETevt, is not included in the horizontal axis of the plot to better demonstrate the main features of the models.}
\end{figure}

\begin{figure*}[th]
\includegraphics[width=\columnwidth]{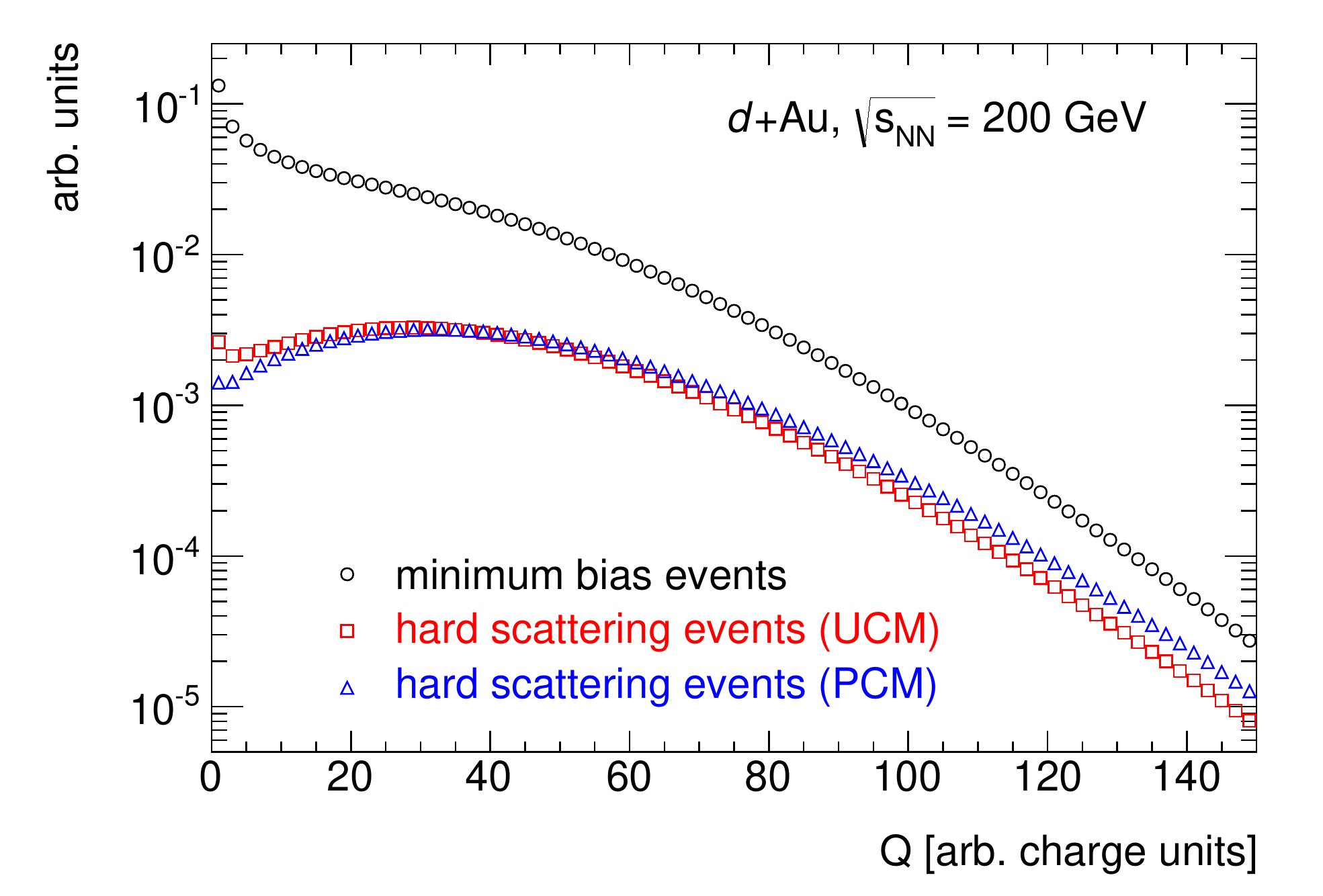}
\includegraphics[width=\columnwidth]{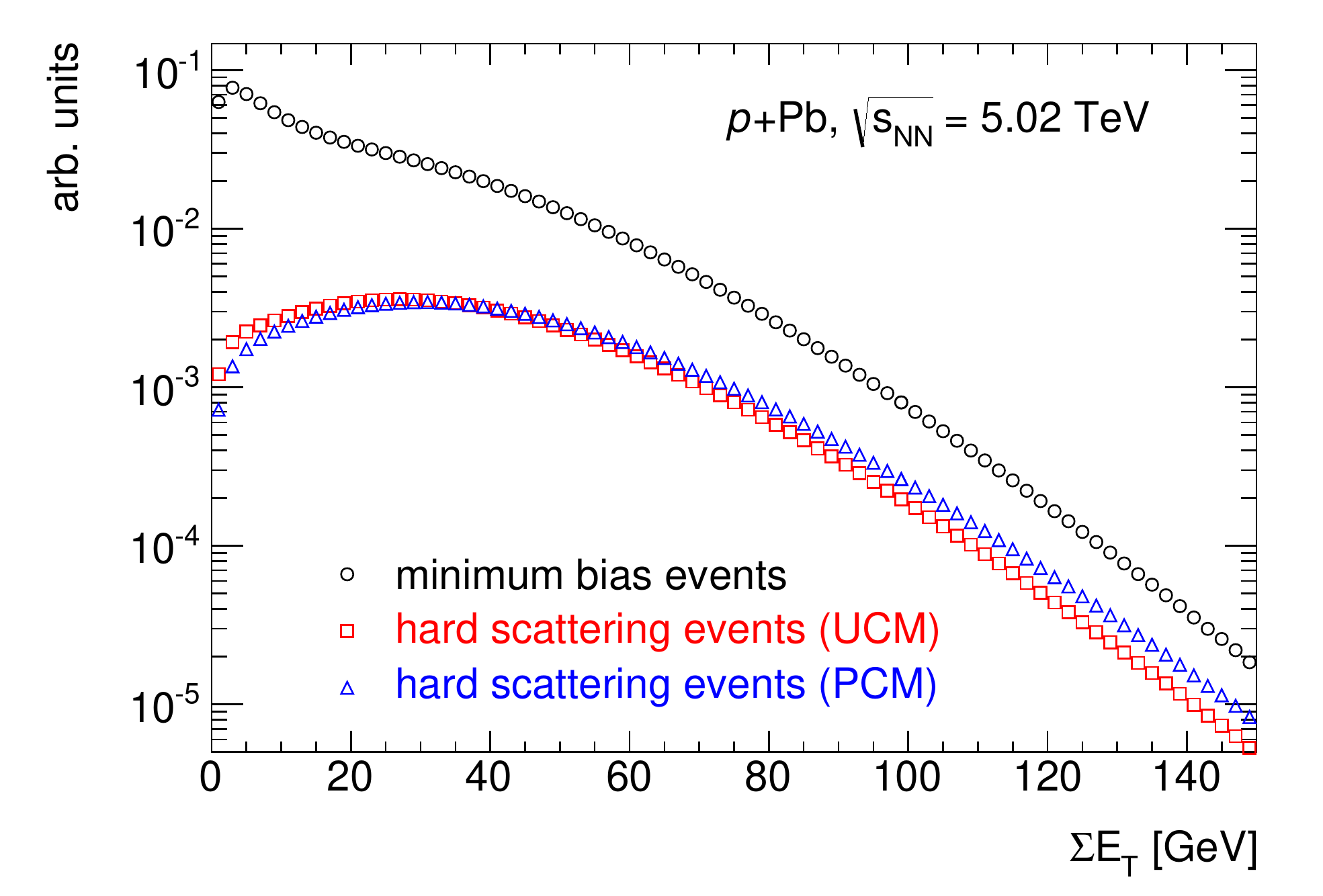}
\caption{\label{fig:yields} Distributions of $Q$ in \dAu\ events (left panel) or \ET\ in \pPb\ events (right panel) for minimum bias events (black circles), compared to the yield-weighted distribution within the UCM (red squares) and the PCM (blue triangles) described in the text, with overall rate $C = 10^{-3}$.}
\end{figure*}

\section{Results} In a given centrality interval, the total hard scattering yield under the assumptions of each model is calculated by integrating the yield from each value of \Ncoll\ within the \ET\ range defining the interval, $(\ETmin, \ETmax)$. Define $\mathcal{Y}$ as the higher-order function which, given a correlation between the average yield and the total \ET, returns the total yield in some centrality interval,

\begin{eqnarray}
 \mathcal{Y}\left[\ Y_\mathrm{N\mathrm{coll}}(\ET); \ETmin, \ETmax\ \right] = \nonumber \\
 \sum_{N\mathrm{coll}} \int_{\ETmin}^{\ETmax}{d\ET} P(\Ncoll) P_{N\mathrm{coll}}(\ET) Y_{N\mathrm{coll}}(\ET) . \label{eq:Yfunctor}
\end{eqnarray}

Let $\rho$ be the ratio of the total yield between one of the correlated models (PCM or VCM) and the UCM in a given centrality interval,

\begin{eqnarray}
\rho = \mathcal{Y}\left[\ Y_\mathrm{N\mathrm{coll}}(\ET)\ \right] /\ \mathcal{Y}\left[\ C \Ncoll\ \right],
\label{eq:nu}
\end{eqnarray}

\noindent where the \ET\ range is suppressed for clarity.

The quantity $\rho$ encodes, in a given experimentally selected centrality interval, how the rate of point-like processes is modified in those events due to the presence of a correlation between the UE activity and the hard scattering yield in \NN\ collisions. Thus, the values of $\rho$ determine how the hard scattering yield in each centrality interval is modified as a result of the centrality bias effect. Experimentally, the values of $\rho$ should be applied to the raw yield $Y$ to determine the corrected yield $Y^\mathrm{corr} = Y / \rho$.

In general, the values of $\rho$ depend on the description of the UE activity (e.g. $k$ and $\theta$) and the chosen centrality intervals. Thus, the prescription presented here must be tailored to the individual conditions within each experiment. This paper gives results for $\rho$ within the \mbox{ATLAS} framework for centrality-selected \pPb\ collisions at $5.02$~TeV, and within the PHENIX framework for \dAu\ collisions at $200$~GeV. 

ATLAS characterizes the centrality of the collision with the sum of the transverse energy, $\ET$, in the Pb-going forward calorimeter~\cite{TheATLAScollaboration:2013cja}. The distribution of $\ET$ is modeled as a Gamma distribution with \Npart-dependent parameters. The best fit uses $k(\Npart) = k_0 + k_1(\Npart-2)$, which is applied in our model by setting $k = k_1 = 0.425$ and $\kevt = k_0 - k_1 = 0.965$. Furthermore, an \Npart-dependent $\theta$ term is used to account for a possible shift of the overall $dN/dy$ distribution of soft particles with increasing \Npart~\cite{Steinberg:2007fg}, which was included in the model via $\theta(\Npart) = 3.41 + 1.30 \log(\Npart-1)$~GeV. The Glauber MC code was used with $\sigmaNN = 70$~mb to generate $P(\Npart)$.  

In PHENIX, centrality is characterized with the total charge $Q$ in the Au-going beam--beam counter~\cite{Adare:2013nff}. The distribution of $Q$ for collisions at a given \Ncoll\ is modeled as an $\Ncoll$-convolution of a negative binomial distribution (NBD) with mean and exponential parameters $\mu = 3.03$ and $\kappa = 0.46$ respectively. Since $\left<Q\right>$ scales linearly with \Ncoll\ (instead of \Npart), the proton contribution term in Eqs.~\ref{eq:ETsum} and~\ref{eq:prob_Ncoll} is neglected in our model. Finally, the normalization in Eqs.~\ref{eq:yield_hardsoft}--\ref{eq:yield_hardsoft4} is modified to use the NBD mean, $1/k\theta \rightarrow 1/\mu$. The Glauber MC code was used with $\sigmaNN = 42$~mb to generate $P(\Npart)$.

In both cases, the centrality divisions were chosen to match those used by the experiments. For each value of \Ncoll\ and \ET\ (in the discussion that follows, respectively $\ET \rightarrow Q$ for the \dAu\ case), the distribution $Y_{N\mathrm{coll}}(\ET)$ was determined by randomly sampling the set of $\left\{ \ETj, \ETpj \right\}$ values and, if necessary, values for \ETevt, the overall contribution to the total \ET\ from the proton participant. This sampling was performed with $10^{5}$ iterations for each value of \Ncoll. 

Fig.~\ref{fig:example_corr} demonstrates the correlation between the hard scattering yield and the centrality signal for \pPb\ events at the LHC with $\Ncoll = 5$ within each model. In the UCM, the mean yield is constant since it is uncorrelated with the centrality signal in any \NN\ collision. In the PCM, the two are generally correlated but with an additional stochastic component arising from the role of \ETpj. Finally, in the VCM with $\alpha=\frac{1}{2}$, where the stochastic component has been integrated out, there is a just positive correlation which agrees with the mean yield at that \ET\ in the PCM.

Fig.~\ref{fig:yields} shows the total $Q$ and \ET\ distributions for minimum bias \dAu\ and \pPb\ events, as well as the hard scattering yield-weighted \ET\ or $Q$ distribution in the UCM and PCM. As expected, the yield-weighted \ET\ or $Q$ distributions in the UCM have a different shape than the minimum bias distribution, reflecting the increase of $\left<\Ncoll\right>$ with increasing \ET. Furthermore, the means of the PCM distributions are shifted to larger \ET\ or $Q$ values than the UCM, reflecting the positive correlation between the \ET\ and average yield in each \NN\ collision. 

\begin{table*}[!t]
\caption{\label{table:results} Multiplicative change in the hard scattering yield, $\rho$, arising from a correlation between the centrality signal and the mean yield in \NN\ collisions. Results are shown for \dAu\ and \pPb\ collisions. The PHENIX bias factors (BF) are taken from the calculation in Ref.~\cite{Adare:2013nff}. For \pPb\ collisions, results are shown for the standard Glauber model and for the Glauber Gribov Color Fluctuation model with two choices of the parameter $\Omega$. }
\begin{center}
\begin{tabularx}{\textwidth}{XXX||XXXX}
\multicolumn{3}{c||}{\dAu\ $200$~GeV} & 
\multicolumn{4}{c}{\pPb\ $5.02$~TeV} \\
\ \ \ \ \ centrality & 
\ \ \ \ \ \ \ \ \ \ $\rho$ & 
1/BF~(PHENIX) & 
\ \ \ \ \ centrality & 
\ \ \ \ $\rho$ (default) & 
\ \ \ $\rho$ ($\Omega=0.55$) & 
\ \ \ $\rho$ ($\Omega=1.01$) \\
\colrule
\multicolumn{1}{c}{}         &  \multicolumn{1}{c}{}               &   \multicolumn{1}{c||}{}              & \multicolumn{1}{c}{0--10\%} & \multicolumn{1}{c}{1.20 $\pm$ 0.10} & \multicolumn{1}{c}{1.09 $\pm$ 0.04} & \multicolumn{1}{c}{1.07 $\pm$ 0.03}\\
\multicolumn{1}{c}{ 0--20\%} & \multicolumn{1}{c}{1.15 $\pm$ 0.07} & \multicolumn{1}{c||}{1.06 $\pm$ 0.01} & \multicolumn{1}{c}{10--20\%} & \multicolumn{1}{c}{1.06 $\pm$ 0.03} & \multicolumn{1}{c}{1.03 $\pm$ 0.02} & \multicolumn{1}{c}{1.03 $\pm$ 0.01} \\ 
\multicolumn{1}{c}{20--40\%} & \multicolumn{1}{c}{0.99 $\pm$ 0.01} & \multicolumn{1}{c||}{1.00 $\pm$ 0.01} & \multicolumn{1}{c}{20--30\%} & \multicolumn{1}{c}{1.00 $\pm$ 0.01} & \multicolumn{1}{c}{1.00 $\pm$ 0.01} & \multicolumn{1}{c}{1.01 $\pm$ 0.01} \\ 
\multicolumn{1}{c}{40--60\%} & \multicolumn{1}{c}{0.92 $\pm$ 0.04} & \multicolumn{1}{c||}{0.97 $\pm$ 0.02} & \multicolumn{1}{c}{30--40\%} & \multicolumn{1}{c}{0.96 $\pm$ 0.02} & \multicolumn{1}{c}{0.98 $\pm$ 0.01} & \multicolumn{1}{c}{0.99 $\pm$ 0.01} \\ 
\multicolumn{1}{c}{60--88\%} & \multicolumn{1}{c}{0.82 $\pm$ 0.09} & \multicolumn{1}{c||}{0.86 $\pm$ 0.06} & \multicolumn{1}{c}{40--60\%} & \multicolumn{1}{c}{0.91 $\pm$ 0.04} & \multicolumn{1}{c}{0.96 $\pm$ 0.02} & \multicolumn{1}{c}{0.97 $\pm$ 0.02} \\
 \multicolumn{1}{c}{}        &  \multicolumn{1}{c}{}               &  \multicolumn{1}{c||}{}               & \multicolumn{1}{c}{60--90\%} & \multicolumn{1}{c}{0.82 $\pm$ 0.07} & \multicolumn{1}{c}{0.87 $\pm$ 0.06} & \multicolumn{1}{c}{0.88 $\pm$ 0.06} \\
 
\end{tabularx}
\end{center}
\end{table*}

Table~\ref{table:results} lists the resulting values of $\rho$ in the PCM for the six \pPb\ and four \dAu\ centrality intervals. Generally, the centrality bias results in an overestimate of the yield in central collisions and an underestimate in peripheral ones. For each system, there is a value of $\rho$ which is nearly unity. For example, the $20$--$30$\% interval in \pPb\ collisions (defined by $31$~GeV $< \ET < 40$~GeV) contains the peak of the yield-weighted \ET\ distribution. Due to this feature, this class of events is relatively insensitive to small shifts in the mean of the distribution, since the increase of the yield on one edge of the interval generally counteracts the decrease at the other.

Intriguingly, the VCM with $\alpha=\frac{1}{2}$ reproduces the results of the PCM to within statistical precision. This is true despite the presence of the stochastic term in Eq.~\ref{eq:yield_hardsoft}, which results in a different description of the \ETj\ vs. $Y_j$ correlation in each \NN\ collision than that in Eq.~\ref{eq:yield_hardsoft2}. This implies that for determining the size of the centrality bias, the details of the correlation in a given \NN\ collision seem to be less important than the overall correlation (assumed here to be linear) between the centrality signal and the average yield, which can be simply parameterized by the factor $\alpha$. 

To demonstrate the sensitivity of $\rho$ to the strength of this overall correlation, different values of $\rho$ were generated by varying $\alpha=0.5$ by 50\% ($\alpha = 0.25$ and $0.75$) in the VCM. This variation represented the midpoint between the default PCM results (with $\alpha=0.5$) and the UCM (with $\alpha=0$). The typical change in $\rho$, shown in Table~\ref{table:results}, is an estimate of the uncertainty in $\rho$ arising from the uncertainty in the strength of the correlation between the hard scattering yield and the UE activity. Additional uncertainties arising from the geometric modeling of \pA\ collisions are not evaluated in this work, but could be determined by evaluating $\rho$ with \Npart\ or \Ncoll\ distributions generated by different sets of MC Glauber parameters as is done in Ref.~\cite{Adare:2013nff}.

Finally, in the \dAu\ and default \pPb\ results, the standard Glauber model with fixed $\sigma_\mathrm{NN}$ was used to determine $P(\Npart)$. However, the procedure is easily generalized to take any $P(\Npart)$ distribution as input, e.g. that provided by the Glauber-Gribov Color Fluctuation (GGCF) model~\cite{Blaettel:1993ah,Alvioli:2013vk}. In this model, which has been investigated by ATLAS~\cite{TheATLAScollaboration:2013cja}, $\sigma_\mathrm{NN}$ varies from event to event to reflect fluctuations in the configuration of the proton wavefunction, resulting in a less steep $P(\Npart)$ distribution at high \Npart\ than in the standard Glauber model. Due to the flatter $P(\Npart)$ distribution, the values of $\rho$ are systematically closer to $1$ than in the default Glauber model. Table~\ref{table:results} lists the values of $\rho$ for the \pPb\ centrality intervals for two choices of the parameter $\Omega$ which characterizes the width of the $\sigma_\mathrm{NN}$ fluctuations.

\section{Discussion} 

The approach presented here was motivated by the approximately direct relationship between the UE activity and the average hard scattering yield observed in \pp\ collisions. Small deviations from the relationship posited in Eq.~\ref{eq:intro}, while straightforward to accommodate in numerical implementations, are not expected to substantially change the values of $\rho$. Nevertheless, additional measurements of how the average yields of hard processes evolve with UE activity in \pp\ collisions would help clarify the picture and refine the results.

Interestingly, although this approach was motivated by studies of the regime in which the mean yield per collision is $C\ll1$, this quantity cancels analytically in Eq.~\ref{eq:nu}. Thus, within the present model, the centrality bias is independent of the overall hard scattering rate and is applicable even if $C>1$.

This approach differs from that proposed by PHENIX~\cite{Adare:2013nff} in several respects. First, the relationship between the hard scattering rate and the charge $Q$ in each \NN\ collision is recast in terms of the average yield as a function of $Q$, instead of a modified $\mathrm{d}N/\mathrm{d}Q$ distribution for a particular \NN\ collision within the \dAu\ collision. Thus the correlation between the yield and $Q$ is treated continuously rather than as a binary division of events into those with and without a hard scattering. Additionally, because the procedure calculates per collision yields instead of probabilities, it naturally incorporates the possibility of multiple hard interactions (or multiple final-state objects) per \pA\ or even \NN\ collision. 

For the \dAu\ system, each value of $\rho$ is compared to the reciprocal of the centrality bias factor (BF) estimated by PHENIX~\cite{Adare:2013nff}. For peripheral events, the BF includes an additional trigger bias correction, which is not modeled in our procedure and thus ignored for the purposes of the comparison. The values of $\rho$ are systematically farther from unity than the corresponding $1/\mathrm{BF}$ values. Nevertheless, the two values are consistent within or only slightly outside the quoted uncertainties in all centrality intervals. Moreover, the sign of the centrality bias effect is the same, such that the bias determined in both models has opposite sign to the high-\pT\ modifications observed in Refs.~\cite{Perepelitsa:2013jua,ATLAS-CONF-2014-024}.

Within the \pPb\ system, ALICE has observed that the apparent strength of the centrality bias in data depends on the pseudorapidity distance, $\Delta\eta$, between the measured hard probe and the centrality detector~\cite{fortheALICE:2013xra,Toia:2014wia}. Thus the VCM with a $\Delta\eta$-dependent $\alpha$ could provide corrections for yields in different kinematic regions. More generally, variations in $\alpha$ could help model the net degree of correlation between the centrality signal and the average yield of hard processes in \NN\ collisions.

\section{Conclusion} 

This paper presents a new approach for determining how the hard scattering yield in centrality-selected \pA\ events is modified from the presence of a positive correlation between the average yield and underlying event activity in \pp\ collisions. Furthermore, it describes how to adapt the procedure to the centrality frameworks used at RHIC and LHC experiments. Finally, the paper discusses implications and possible extensions of this procedure.


\begin{acknowledgements} The authors thank Brian Cole for insightful discussions.
\end{acknowledgements}

\bibliography{centbiaspaper}

\end{document}